\documentclass[pra,twocolumn,showpacs,superscriptaddress]{revtex4-1}

\usepackage{graphicx}
\usepackage{multirow}
\usepackage{rotating}
\usepackage{wasysym}

\begin{document}

\title{Fulde-Ferrell-Larkin-Ovchinnikov critical polarization in 1D fermionic optical lattices}

\author{Vivian V. Fran\c{c}a}
\affiliation{Physikalisches Institut der Albert-Ludwigs Universit\"at, Hermann-Herder-Str.3, Freiburg, Germany}
\affiliation{Capes Foundation, Ministry of Education of Brazil, Caixa Postal 250, Brasilia, 70040-20, Brazil}

\author{Dominik H\"orndlein}
\affiliation{Physikalisches Institut der Albert-Ludwigs Universit\"at, Hermann-Herder-Str.3, Freiburg, Germany}

\author{Andreas Buchleitner}
\affiliation{Physikalisches Institut der Albert-Ludwigs Universit\"at, Hermann-Herder-Str.3, Freiburg, Germany}
%\date{\today}

\begin{abstract}
We deduce an expression for the critical polarization $P_C$ below which the FFLO-state emerges in one-dimensional lattices with spin-imbalanced populations. We provide and explore the phase diagram of unconfined chains as a function of polarization, interaction and particle density. For harmonically confined systems we supply a quantitative mapping which allows to apply our phase diagram also for confined chains. We find analytically, and confirm numerically, that the upper bound for the critical polarization is universal: $P_C^{max}=1/3$ for any density, interaction and confinement strength.\end{abstract} 


\newcommand{\be}{\begin{equation}}
\newcommand{\ee}{\end{equation}}
\newcommand{\bea}{\begin{eqnarray}}
\newcommand{\eea}{\end{eqnarray}}
\newcommand{\bi}{\bibitem}
\newcommand{\la}{\langle}
\newcommand{\ra}{\rangle}
\newcommand{\ua}{\uparrow}
\newcommand{\da}{\downarrow}
\renewcommand{\r}{({\bf r})}
\newcommand{\rp}{({\bf r'})}
\newcommand{\eps}{\epsilon}
\newcommand{\bfr}{{\bf r}}

\maketitle

Superconductivity, which has celebrated 100 years \cite{5}, has fascinated scientists since it was discovered \cite{6}. 
In non-polarized, fermionic systems, superfluidity is described by Bardeen-Cooper-Schrieffer (BCS) theory \cite{7}, 
in which attractive fermions with zero-momentum are paired. 
External magnetic fields, as well as polarization induced by spin-imbalanced populations, are expected to destroy 
the BCS-pairing mechanism and force the system into a polarized normal Fermi liquid.
However, according to Fulde and Ferrell \cite{1} and, independently, to Larkin and Ovchinnikov \cite{2}, there would exist 
a polarization regime for which superfluidity still survives against the normal regime, referred to as FFLO phase. 
This exotic coexistence of superconductivity and magnetism is predicted to manifest by a spontaneous breaking of spatial symmetry, characterized by an oscillating order parameter \cite{10}. 

Experimentally, there are only indirect evidences of the FFLO phase, in solid-state materials \cite{solid} and in 1D tubes \cite{3}. Theoretically, considerable progress has been achieved in understanding the FFLO general properties \cite{10, 15, 20, 21, 22, 23, 24, 25, 26, 27, 28},
regarding the complexity of dealing with many-particle interactions, large systems, and the harmonic confinement necessary to 
describe state-of-the-art experiments. The regime of polarizations at which the FFLO-phase can be found has been determined however only on empirical grounds: for specific systems and parameters, by means of FFLO witnesses. 

We here derive an expression from general, system-independent considerations for the critical polarization $P_C$ below which the FFLO-state emerges.  We depict the phase diagram for unconfined systems and provide a quantitative mapping which allows its application also to harmonically confined chains. Although the critical polarization exhibits a strong parameter dependence, we find that its upper bound is universal:  $P_C^{max}=1/3$, for any density, interaction and confinement strength. 

Our method consists in solving the one-dimensional Hubbard model \cite{16},

\begin{equation}
\hat{H}=-t\sum_{<ij>,\sigma}\hat{c}^\dagger_{i\sigma}\hat{c}_{j\sigma} + U\sum_i \hat{n}_{i\uparrow}\hat{n}_{i\downarrow}+ V\sum_{i\sigma} x_i^2 \hat{n}_{i\sigma},\label{hm}
\end{equation}
at zero temperature, with onsite attractive interaction $U$, next-neighbor tunneling $t$, harmonic confinement of strength $V$, 
lattice size $L$, total number of particles  $N=N_\uparrow+N_\downarrow$, with density $n=N/L=n_\uparrow+n_\downarrow$ 
and spin-imbalance $P=(N_\uparrow - N_\downarrow)/N$. Furthermore, $\hat{c}^\dagger_{i\sigma}$ ($\hat{c}_{i\sigma}$) is the fermionic creation (annihilation) operator, $\hat{n}_{i\sigma}=\hat{c}^\dagger_{i\sigma}\hat{c}_{i\sigma}$ counts the particle number on site $i$, and $\sigma$ runs over the spin values $\uparrow$ and $\downarrow$. Ground-state properties of finite chains are obtained via Density-Matrix Renormalization Group (DMRG) \cite{dmrg} and Density Functional Theory (DFT) \cite{18} calculations, the latter with a local spin-density approximation to the exchange-correlation energy \cite{19}. A Bethe-Ansatz (BA) solution is used for average properties of infinite chains, through a recently derived analytical parametrization (BA-FVC) \cite{17}.

Let us first consider the spin state occupation probabilities. Each chain site has a four-dimensional Hilbert space: paired spins, single occupation of a spin-up or spin-down particle, and vacuum. The key quantities are the paired probability, 
\begin{eqnarray}
\text{w}_{\uparrow\downarrow}\equiv\frac{1}{L}\sum_i\left\langle \hat{n}_{i\uparrow}\hat{n}_{i\downarrow}\right\rangle=\frac{1}{L}\frac{\partial E_0(n,P,U)}{\partial U},\label{w2} 
\end{eqnarray}
closely related to superfluidity, and the majority unpaired probability, 
\begin{eqnarray}
\text{w}_\uparrow\equiv\frac{1}{L}\sum_i\left\langle 
\hat{n}_{i\uparrow}\right\rangle-\text{w}_{\uparrow\downarrow}
=\frac{n}{2}\left[1+P\right]-\text{w}_{\uparrow\downarrow}, \label{wup}
\end{eqnarray}
associated with the normal phase, where $E_0(n,P,U)$ is the ground-state energy.

We start, in Figure 1, by monitoring $\text{w}_{\uparrow\downarrow}$ and $\text{w}_\uparrow$ as a function of the spin-imbalance 
in the strongly interacting regime ($U=-8t$). The probabilities are found to be independent on the confinement in this regime, what is plausible, since the harmonic confinement plays a less important role for strong interactions. Also, as $\text{w}_{\uparrow\downarrow}$ and $\text{w}_\uparrow$ are average quantities, we observe that the usually less precise methods, BA-FVC for unconfined and DFT for confined systems, show very good agreement with the more precise DMRG data. In particular, for the intersection point $P_I\equiv P(\text{w}_\uparrow=\text{w}_{\uparrow\downarrow})$, Table 1 shows that this good agreement holds for any density and interaction strength. The rather good performance of BA-FVC and DFT allows us to obtain $P_I$ analytically for unconfined chains (via BA-FVC) and numerically, at about $1 \permil$ of the computing time of the DMRG calculation, for confined systems (via DFT).

\begin{figure}
\hspace{-0cm}\includegraphics[width=8.5cm]{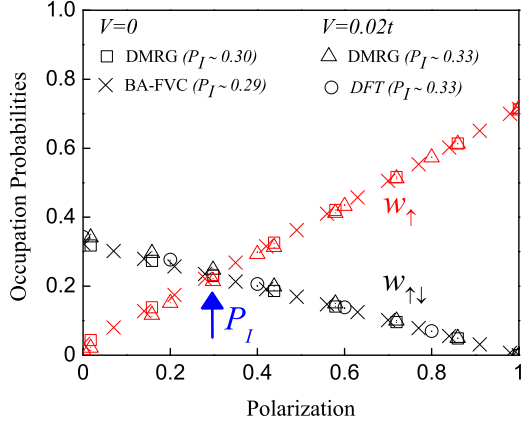}\vspace{-0cm}
\caption{Occupation probabilities (unpaired $\text{w}_\uparrow$ and 
paired $\text{w}_{\uparrow\downarrow}$), averaged over the entire lattice, as a function of polarization $P$. The intersection point 
$P_I$ coincides for different computational approaches, for confined ($V=0.02t$) and unconfined chains ($V=0$), in the strongly interacting regime, $U=-8t$, and at particle density $n=0.7$: analytical BA-FVC results 
for the infinite chain, DFT and DMRG calculations for finite chains ($L=80$)  with open boundary conditions. }
\label{fig1}
\end{figure}

\begin{table}[b]
\centering
\caption{Critical Polarization $P_C$ for several densities and interaction regimes, confined and unconfined chains, obtained with DMRG and DFT, for chains with open boundary conditions and $L=80$ sites, and by BA-FVC \cite{17}, for infinite chains.} \label{tab}
\vspace{0.2cm}
\begin{tabular}{  cc|cc|cc}
&&\multicolumn{2}{c|}{Unconfined $V=0$}&\multicolumn{2}{c}{Confined $V=0.02t$}\\
&n& DMRG & BA-FVC& DMRG & DFT\\
\hline
\multirow{4}{*}{U=-8t}&0.2& 0.28 &0.30&0.32 &0.32\\
&0.5&0.29 &0.29 &0.33 & 0.33 \\
&0.7&0.29 &0.29 &0.33 & 0.33 \\
&1.0&0.31 &0.31 &0.33 & 0.33 \\
\hline
\multirow{4}{*}{U=-4t}&0.2& 0.17 &0.13 &0.27 &0.26   \\
&\normalsize 0.5&0.19 &0.15 & 0.32& 0.33 \\
&0.7&0.21 &0.20 & 0.32 & 0.33 \\
&1.0&0.27 &0.26 &0.33 & 0.33 \\
\hline
\multirow{4}{*}{U=-2t}&0.2& 0.00 &0.00 &0.17 &0.14  \\
&\normalsize 0.5&0.00 &0.00 & 0.30& 0.33 \\
&0.7&0.07 &0.00 &0.30& 0.33 \\
&1.0&0.19 &0.18 & 0.33& 0.33 \\
\end{tabular}
\end{table}

\begin{figure}
\hspace{-0cm}\includegraphics[width=8cm]{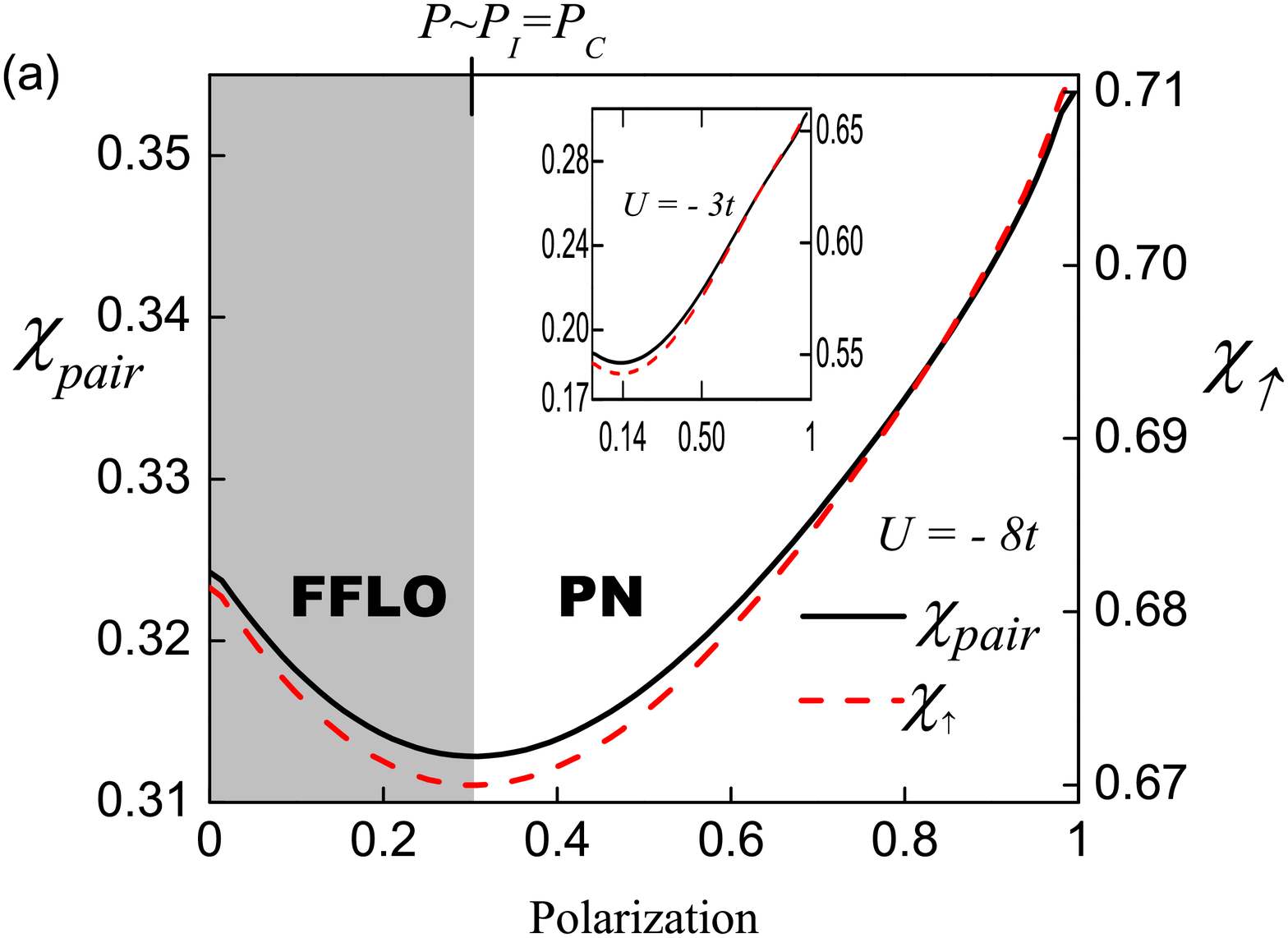}
\hspace{-0cm}\includegraphics[width=8cm]{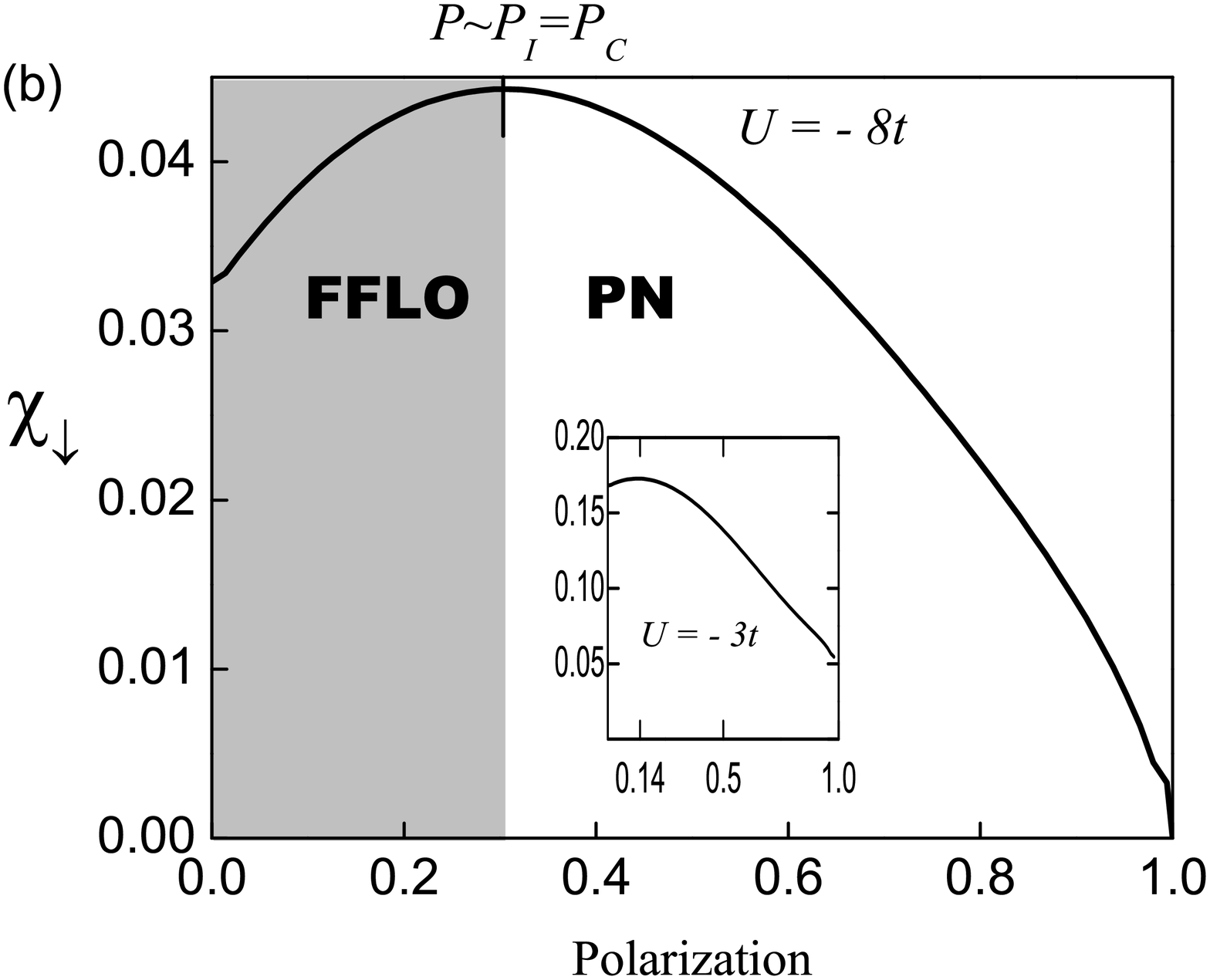}
\caption{a) Pair and majority unpaired susceptibilities and b) minority species susceptibility,  for $n=0.7$ and $U=-8t$, extracted from the analytical BA-FVC results, which clearly exhibit an inflection 
point at $P\approx P_I$. This corresponds to the critical polarization $P_C$ which delimits the superfluid phase, for $P<P_C$ (FFLO), 
and the polarized normal phase, for $P>P_C$ (PN). The insets show the same quantities for $U=-3t$: as the pairs are less bound, $\chi_{pair}$ varies in a larger range (here $P_C(n=0.7, U=-3t)\approx 0.14$).}
\label{fig1}
\end{figure}

At a first glance, Fig.1 suggests no special feature for the probabilities in the entire range of polarization: they seem to 
vary linearly with $P$. If this was the case, the pair susceptibility $\chi_{pair}=|d\text{w}_{\uparrow\downarrow}/dP|$ 
(or equivalently, the majority species susceptibility $\chi_{\uparrow}=|d\text{w}_{\uparrow}/dP|$) would be constant in $P$, what means 
that there would exist no special regime of polarization in which the pairing mechanism, or equivalently superfluidity, is protected. 
Figure 1b instead reveals that $\chi_{pair}$ is {\it not} a constant: On the contrary, it clearly shows two distinct phases with inflection 
point precisely at the intersection point $P_I$. For $P<P_I$ the pair susceptibility decreases with imbalance, i.e., the system acts {\it against} the increase of $P$, to protect the pairing mechanism $-$ what is consistent with the FFLO 
superconducting phase predicted behavior. On the other hand, for $P>P_I$, the unprotected pairs are increasingly more 
susceptible to $P$, i.e., the system acts in {\it favor} of enhanced polarization $-$ characteristic of a normal, 
non-superfluid phase \cite{10}. Similar behavior is also observed for weaker interactions, as shows the inset of Fig.1b for $U=-3t$. 

The pairing mechanism in the FFLO regime is protected by the unpaired minority species, $\text{w}_{\downarrow}$ (for $P>0$). When increasing the imbalance (for constant total number of particles), the system has two flipping channels: either (I) from an unpaired state, $\left|\downarrow\right\rangle\rightarrow\left|\uparrow\right\rangle$, or (II) from a paired state, $\left|\uparrow\downarrow\right\rangle\rightarrow\left|\uparrow\right\rangle,\left|\uparrow\right\rangle$. Energetically, channel (I) is favored, with energy cost restricted to the polarization, while in (II) there is the additional cost of breaking a pair. So one would expect that channel (I) is preferable for any $P$. However, since the probability of finding the state $\left|\downarrow\right\rangle$ is very small (though finite) for all $P$ (due to the attractive interaction), spin flips always have contributions from both channels (I) and (II).  Note that channel (I) does not affect pairing, while channel (II) does. Consequently, the larger the weight of (I), determined by the susceptibility $\chi_\downarrow=|d\text{w}_\downarrow/dP|$, the more robust are the pairs. This is supported by Fig.2b: For $P<P_I$, $\chi_\downarrow$ {\it increases} with polarization, while for $P>P_I$ it {\it decreases} with $P$ increasing, due to the minority species' then almost negligible population. Beyond $P_C$, polarization is therefore progressively created by the breaking of pairs, channel (II), and the superfluid component must fade away.

\begin{figure}[t]
\hspace{-0.4cm}\includegraphics[width=9cm]{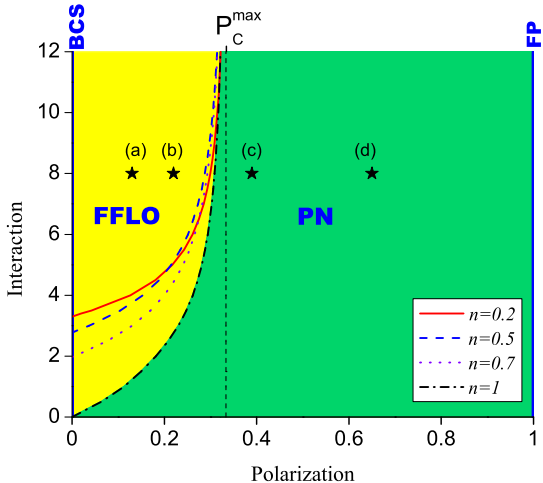}\hspace{-0cm}\vspace{-0.2cm}
\caption{
Phase diagram as a function of polarization $P$ and attractive interaction with strength $|U|/t$, for unconfined chains (for $V\neq 0$ the demarcation line between FFLO and PN is shifted according to Fig.5). Yellow (light gray) and green 
(dark gray) areas represent superfluid (FFLO) and polarized normal (PN) fluid, respectively. $P=0$ identifies the BCS phase, 
while $P=1$ the fully polarized phase (FP). The demarcation line between FFLO and PN, obtained via Eq.(\ref{pc}), is shown for four 
densities. The vertical dashed line delimits the analytically found upper bound $P_C^{max}=1/3$. Stars represent the situations depicted in Fig.4, panels (a)$-$(d).  }
\label{fig2}
\end{figure} 

\begin{figure}
\vspace{-0.3cm}\hspace{-0.9cm}\includegraphics[width=5cm]{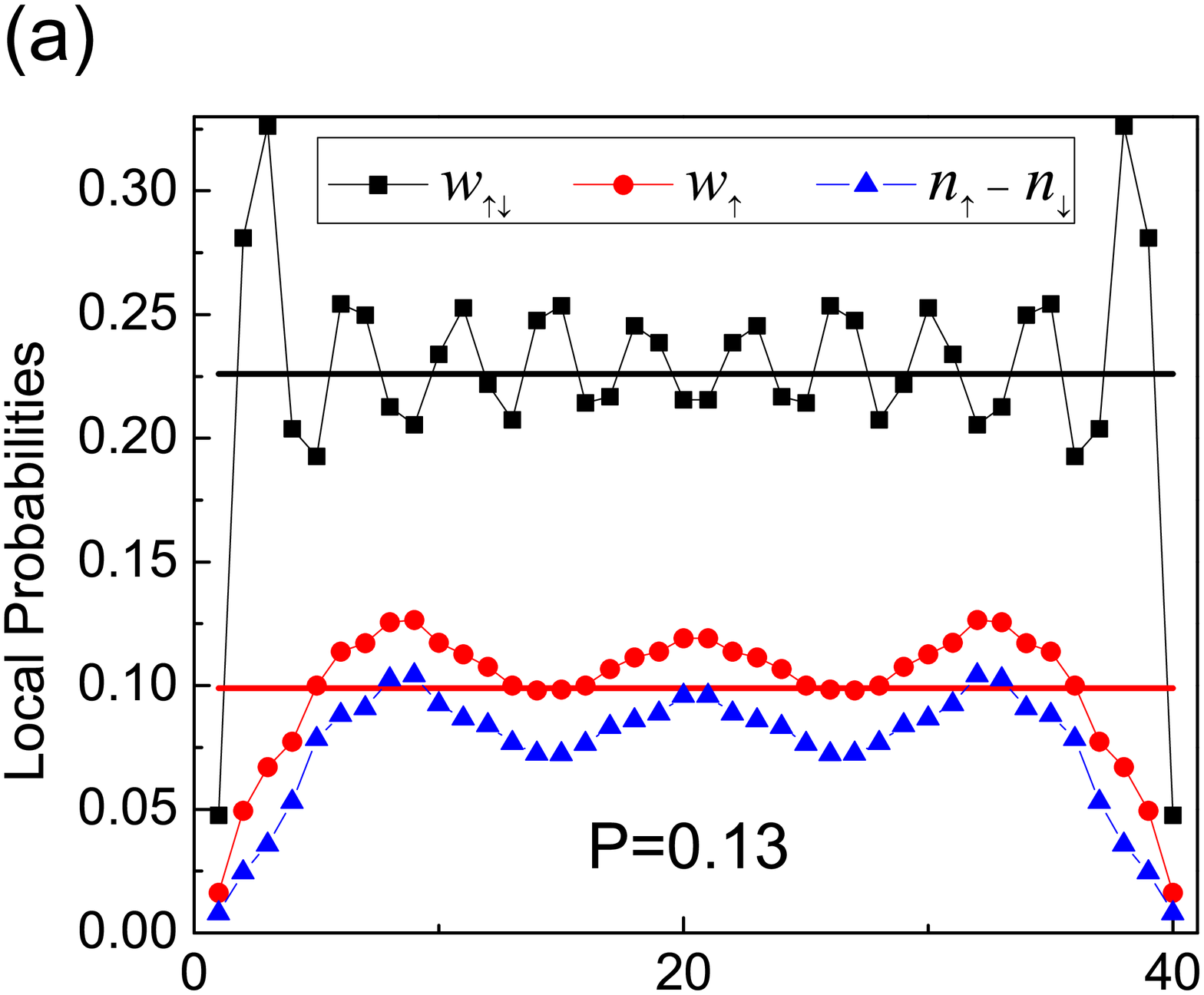}\hspace{-0.7cm}
\includegraphics[width=5cm]{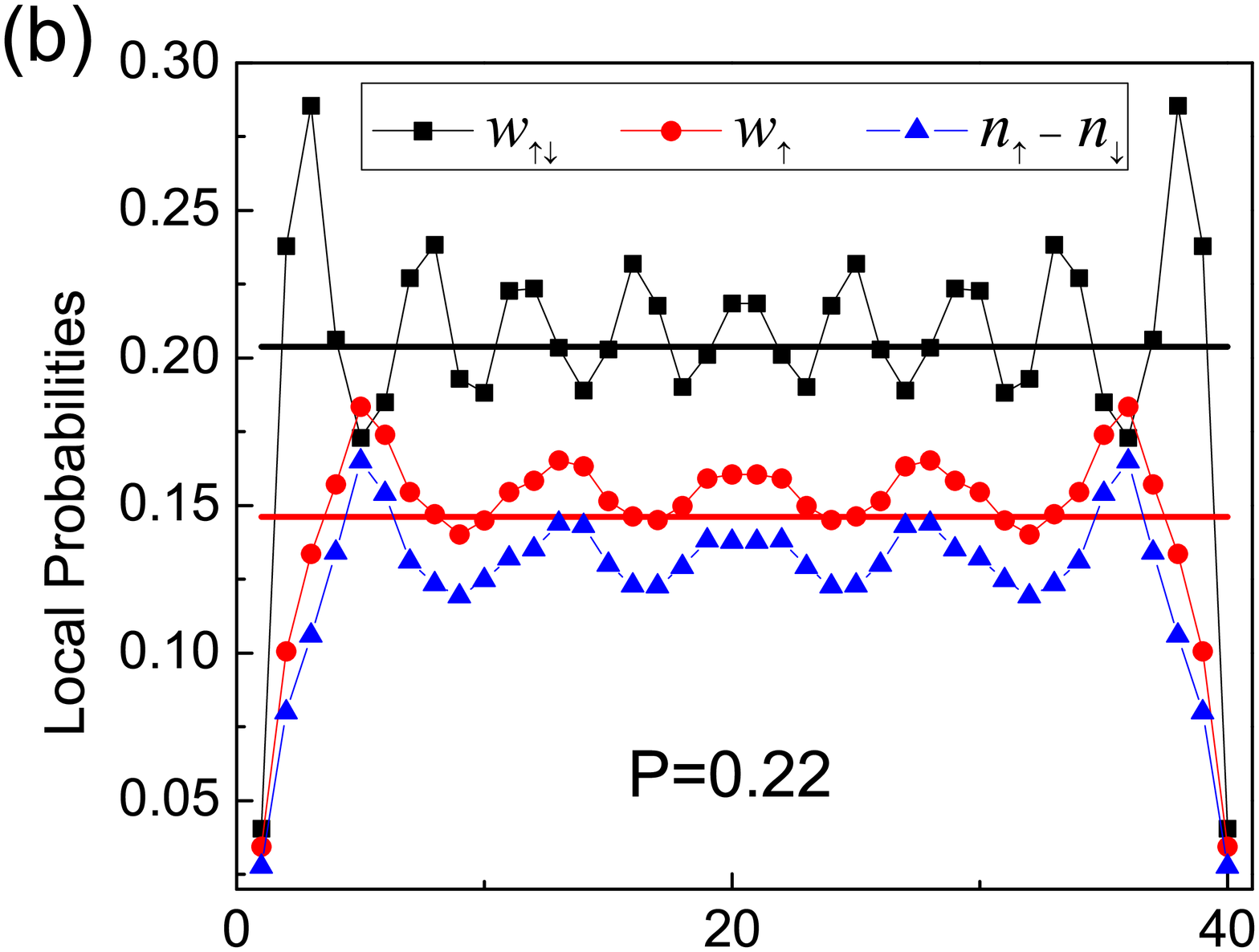}\hspace{-1.2cm}\vspace{-0.5cm}

\hspace{-0.9cm}\includegraphics[width=5cm]{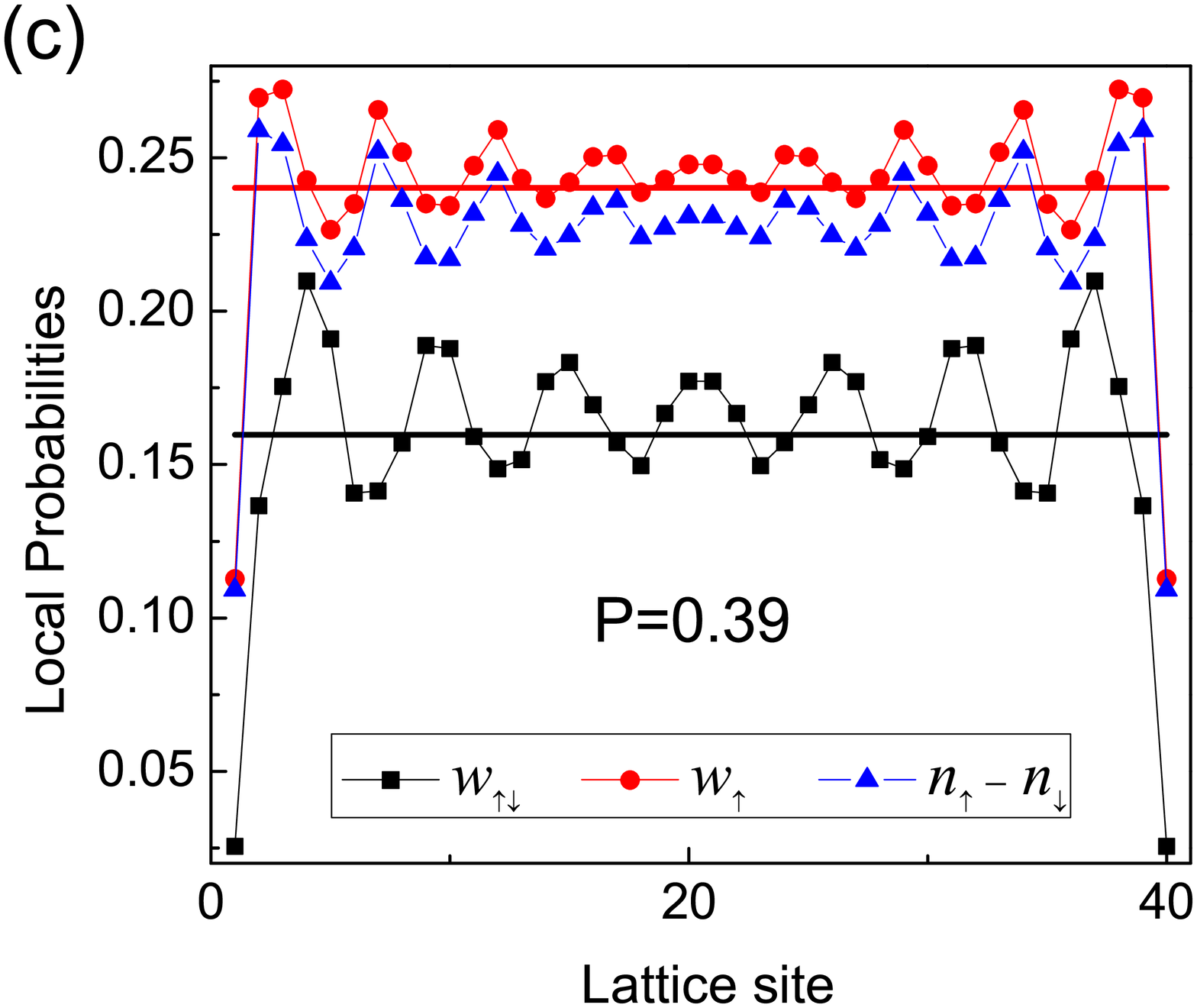}\hspace{-0.7cm}
\includegraphics[width=5cm]{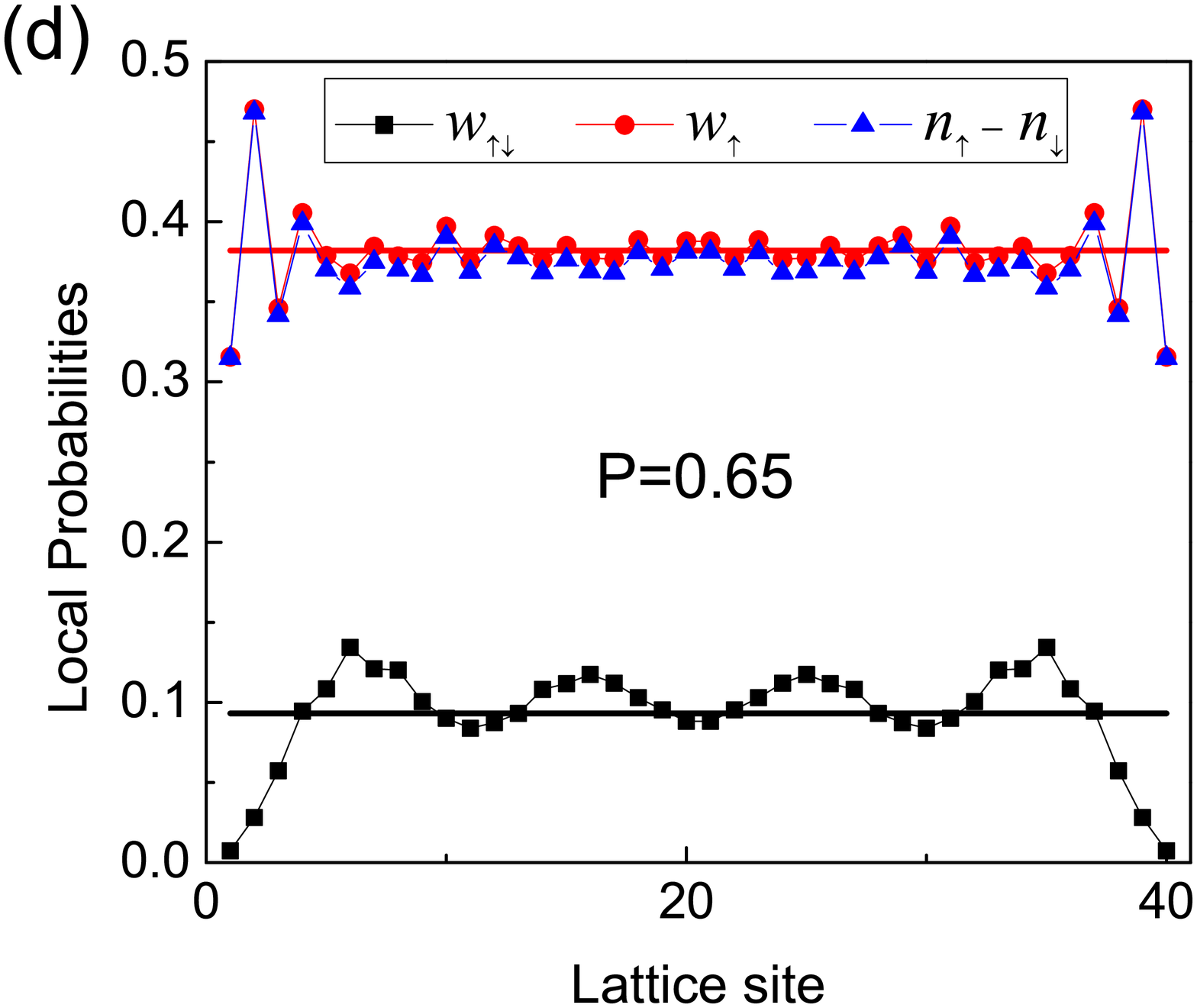}\hspace{-1.2cm}
\caption{
Evolution of magnetization ($m=n_\uparrow-n_\downarrow$), unpaired ($\text{w}_{\uparrow}$) and paired ($\text{w}_{\uparrow\downarrow}$) probability profiles with imbalance $P$: (a) and (b)
in the FFLO regime ($P<P_C$); (c) and (d) for a polarized normal phase ($P>P_C$). Solid lines indicate the averages of $\text{w}_{\uparrow\downarrow}$ and $\text{w}_{\uparrow}$  over the entire lattice. The profiles are obtained via DMRG for $L=40$, $N=23$, $U=-8t$ and $V=0$. $P_C(n=0.58,U=-8t)=0.29$, and each panel is located $-$ by an asterisk $-$ in the phase diagram in Fig.3.}
\end{figure}

Thus $P_I$ has an important physical meaning: it corresponds to the critical polarization $P_C$ below which the FFLO-state emerges, 
and is given by

\begin{equation}
P_C(n,U)=\pm \left[\frac{4\text{w}_{\uparrow\downarrow}(n,P_C,U)}{n}-1\right],\label{pc}
\end{equation}
where we have set $\text{w}_{\uparrow\downarrow}=\text{w}_{\uparrow}$ in equations (\ref{w2})$-$(\ref{wup}). The sign in Eq.(\ref{pc}) depends on the majority species, spin up (+) or down ($-$). Figure 3 shows the phase diagram as a function of interaction and polarization, with the demarcation 
line between FFLO and polarized normal phase (PN) (defined by Eq.(\ref{pc})) indicated for various densities $n$. We find that for small $U$ and $n$ the FFLO-phase is suppressed: already small, finite values of $P$ suffice to induce a transition directly from the BCS to a partially polarized normal regime \cite{20}. For larger 
densities and weak interactions, the FFLO area increases with $n$, whereas it is nearly independent of $n$ for larger values of $U$. That because the larger the density the larger the initial (at  $P=0$) pairing probability, which helps the superfluidity to prevail for higher polarizations at moderate $U$, but becomes irrelevant for strong interactions.

In fact, one can analytically obtain this upper bound for the critical polarization, $P_C^{max}$, by calculating the maximum value of $w_{\uparrow\downarrow}$ (see Eq.(4)). By applying usual particle-hole transformations between attractive and repulsive systems \cite{31,32}, 
\begin{equation}
\text{w}_{\uparrow\downarrow}(n,P,U<0)=\frac{n}{2}\left[1-|P|\right]-\text{w}_{\uparrow\downarrow}(n',P',|U|),
\label{1.9}
\end{equation}
where $n'=nP+1$ and $P'=(n-1)/n'$, we see that the maximal value $\text{w}^{max}_{\uparrow\downarrow}(U<0)$ corresponds to the minimum of $\text{w}_{\uparrow\downarrow}(|U|)$. This is precisely the case at $U\rightarrow -\infty$: double occupancy in the repulsive system vanishes, $\text{w}_{\uparrow\downarrow}(|U|\rightarrow \infty)=0$, and double occupancy in the attractive system, Eq. (5), reaches its maximum, $\text{w}_{\uparrow\downarrow}^{max}(U\rightarrow -\infty)=(n/2)(1-|P|)$. Plugging this latter in our formula (4) for $P_C$, we find the upper bound
\begin{equation}
\left|P_C^{max}\right|=\frac{1}{3},
\label{2.2}
\end{equation}
which is completely consistent with the phase diagram (Fig.3) and with our numerical observations for large $U$ (Table 1).

\begin{figure}
\vspace{-0.1cm}\includegraphics[width=8.5cm]{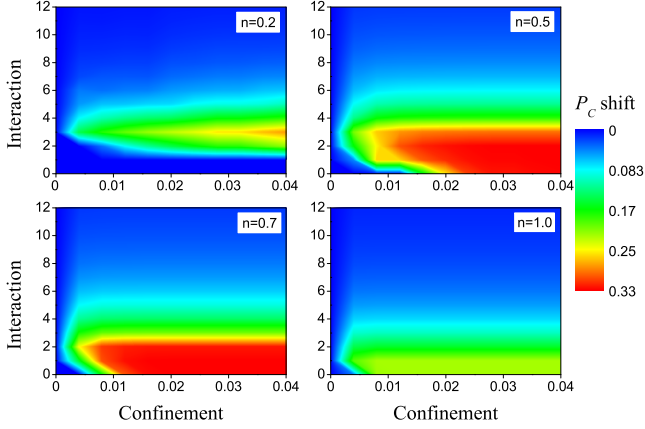}\vspace{-0.4cm}
\caption{
Mapping between confined and unconfined chains via DFT calculations: Shift of the critical polarization, $P_C(n,U,V)- P_C(n,U,V=0)$, as a function of interaction $|U|/t$ and confinement 
strength $V/t$, for different particle densities. $L=N=80$ and $U=-8t$.}
\label{fig3}
\end{figure}

In the limit of very strong attractive interactions, $U\rightarrow -\infty$, Eq.(\ref{1.9}) reads
\begin{equation}
\text{w}_{\uparrow\downarrow}(n,P,U\rightarrow -\infty)=\frac{n}{2}\left[1-|P|\right],
\label{1.10}
\end{equation}
since in this limit the double occupancy for the repulsive system vanishes, $\text{w}_{\uparrow\downarrow}(n',P',|U|\rightarrow \infty)=0$. Thus, the majority-spin probability for $P>0$ (Eq.(3)) is given by 
\begin{eqnarray}
\text{w}_\uparrow(n,P,U\rightarrow -\infty)&=&nP\nonumber\\
&=&\frac{N}{L}\left(\frac{N_\uparrow-N_\downarrow}{N}\right)\nonumber\\
&=&n_\uparrow-n_\downarrow,
\label{1.11}
\end{eqnarray}
where $m=n_\uparrow-n_\downarrow$ is the magnetization. This convergence of the averages $\text{w}_\uparrow$ and $m$ in the limit $U\rightarrow -\infty$ suggests that, for strong but finite interactions, the 
general properties of the unpaired probability profile can be observed in the local magnetization. Indeed, in Figure 4, we verify numerically the {\it local} agreement between $\text{w}_{i,\uparrow}$ and $m_i$, for $U=-8t$. In the superfluid regime, $P<P_C$ (Figs.4a, 4b), we find that $\text{w}_{\uparrow\downarrow}$ dominates the entire chain, while the unpaired spins are inhomogeneously distributed, with an accumulation (local probability above the average) at the center of the chain. In the normal regime, $P>P_C$ (Figs.4c, 4d), the opposite occurs: $\text{w}_{\uparrow}$ prevails over the pairs in the entire chain, with Friedel-like oscillations around its average, while $\text{w}_{\uparrow\downarrow}$ accumulates at the chain center. 

\begin{figure}
\vspace{-0.1cm}\includegraphics[width=7cm]{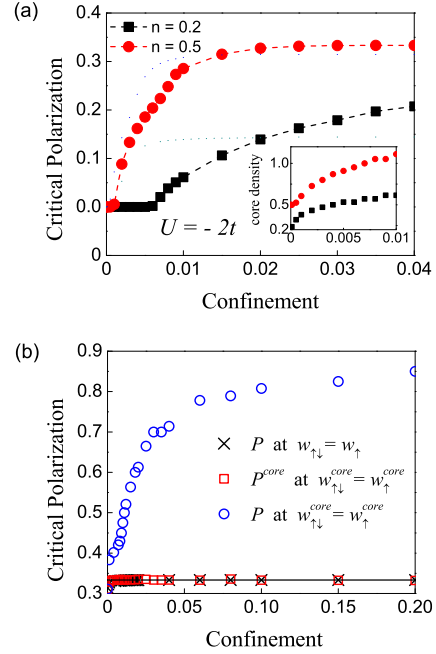}\vspace{-0.0cm}
\caption{
a) Critical polarization $P_C$ as a function of the confinement strength $V$, in the regime of weak interactions and small densities 
where, at $V=0$, there is no FFLO-phase ($P_C=0$, also see Table I and Fig. 3). As $V$ increases, the FFLO-state emerges (finite $P_C$). Inset: 
The effective density at the potential core, as defined in Ref.\cite{22}, in units of the average density $n$, as a function of $V$. b) Critical polarization of confined chains, evaluated: 
i) for the entire chain ($P$ at $\text{w}_{\uparrow\downarrow}=\text{w}_{\uparrow}$), ii) at the core ($P^{core}$ at $\text{w}^{core}_{\uparrow\downarrow}=\text{w}^{core}_{\uparrow}$) 
and iii) for the entire chain, though at the onset of the normal phase in the core ($P$ at $\text{w}^{core}_{\uparrow\downarrow}=\text{w}^{core}_{\uparrow}$). 
The solid line indicates the analytical upper bound $P_C^{max}=1/3$.
Here $L=N=80$ and $U=-8t$.}
\label{fig3}
\end{figure}

Experimentally, ultracold atoms are among the most suitable systems for the detection of exotic superfluidity \cite{3,10,13,14}. 
For strongly interacting lattice systems ($|U|\geq 8t$), we show, however, in Fig.1 and Table 1, that the experimental \cite{12} harmonic 
confinement $V$ only slightly shifts the critical polarization. Therefore, the phase diagram in Fig.3, constructed for 
unconfined chains, can be directly applied for harmonically confined systems in this regime of interaction. For weakly interacting 
systems, though, depending on the actual particle density, $V$ implies a significant shift of $P_C$ which cannot be neglected anymore. 

Hence, by comparing the critical polarizations of confined and unconfined chains, obtained with DFT 
(similar to Fig.1, but now for several $n$ and $U$), we provide, in Figure 5, the necessary shift on $P_C$, which allows one 
to apply our phase diagram for any harmonically confined system. Independently of the required shift, we find that the critical 
polarization is still limited to $P_C^{max}=1/3$ for any density, interaction and confinement strength, i.e., the analytical upper bound 
obtained here is {\it universal}. 

Finally, we remark that this upper bound, which is consistent with the original prediction and also with quantum Monte Carlo calculations \cite{28}, has an apparent discrepancy with recent investigations in harmonically confined systems \cite{21, 22, 23, 24, 25, 26, 27}, which report FFLO at higher polarizations, as much as $P_C\approx0.8$. This is however a simple artifact of the confinement: While unconfined chains present no FFLO-state ($P_C=0$) for small $n$ and $U$, Figure 6a reveals that, depending on the intensity $V$, harmonically confined systems, conversely, do have an FFLO phase ($P_C\neq0$) (see also DMRG data for U=-2t in Table 1). 
That because $V$ reduces the volume where both species can be found (the central core) \cite{21,22}, 
such that the core density actually is much larger than the average $n$ (see inset of Fig.6a) and thus the FFLO-state emerges. 
This smaller effective volume is also responsible for the observations of FFLO at polarizations beyond $P_C^{max}$ \cite{21, 22, 23, 24, 25, 26, 27}: 
while the FFLO witnesses were observed exclusively at the central core, they were associated with a larger {\it global} polarization 
(defined for the entire chain), resulting in values of $P_C$ much larger than our upper bound $P_C^{max}=1/3$. This apparent discrepancy disappears though whenever one compares the FFLO signatures (in our case $\text{w}_{\uparrow\downarrow}^{core}$ and 
$\text{w}_{\uparrow}^{core}$) to the polarization within the core, as we do in Figure 6b.

We thank Peter Fulde, Randall Hulet and Fabian Heidrich-Meisner for fruitful discussions, and Michael Lubasch and Andr\'e Malvezzi for the exchange and discussion of benchmark data on the ground state energies and electron distributions of extended Fermi Hubbard chains. V.F. is indebted to CAPES (4101-09-0) and to the {\it F\"orderung evaluierter Forschungsprojekte} of the University Freiburg.

\end{document}